# On the Nature of the Dust in the Debris Disk Around HD69830

C.M. Lisse[1], C.A. Beichman[2,3], G. Bryden[3], M.C. Wyatt[4]

[1] Planetary Exploration Group, Space Department, Johns Hopkins University Applied Physics Laboratory, 11100 Johns Hopkins Rd, Laurel, MD 20723   carey.lisse@jhuapl.edu

[2] Michelson Science Center, California Institute of Technology, M/S 100-22, Pasadena, CA 91125   chas@pop.jpl.nasa.gov

[3] Jet Propulsion Laboratory, 4800 Oak Grove Drive, Pasadena, CA 91109   Geoffrey.Bryden@jpl.nasa.gov

[4] Institute of Astronomy, University of Cambridge, Madingley Road, Cambridge CB3 0HA, UK   wyatt@ast.cam.ac.uk

Proposed Running Title: **Dust in the Debris Disk Around HD69830**

Please address all future correspondence, reviews, proofs, etc. to :

Dr. Carey M. Lisse

Planetary Exploration Group, Space Department

Johns Hopkins University, Applied Physics Laboratory

11100 Johns Hopkins Rd

Laurel, MD 20723

240-228-0535 (office) / 240-228-8939 (fax)

Carey.Lisse@jhuapl.edu

# Abstract

We have used the infrared mineralogical model derived from the Spitzer IRS observations of the Deep Impact experiment to study the nature of the dust in the debris found around the K0V star HD69830 (Beichman *et al.* 2005). Using a robust approach to determine the bulk average mineralogical composition of the dust, we show it to be substantially different from that found for comet 9P/Tempel 1 (Lisse *et al.* 2006) or C/Hale-Bopp 1995 O1 and comet-dominated YSO HD100546 (Lisse *et al.* 2007). Lacking in carbonaceous and ferrous materials but including small icy grains, the composition of the HD 69830 dust resembles that of a disrupted P or D-type asteroid. The amount of mass responsible for the observed emission is the equivalent of a 30 km radius, 2500 kg m$^{-3}$ sphere, while the radiative temperature of the dust implies that the bulk of the observed material is at ~1.0 AU from the central source, coincident with the 2:1 and 5:2 mean motion resonances of the outermost of 3 Neptune-sized planets detected by Lovis *et al.* (2006). In our solar system, P and D-type asteroids are both large and numerous in the outer main belt and near Jupiter (e.g. the Hildas and Trojans) and have undergone major disruptive events to produce debris disk-like structures (c.f. the Karin and Veritas families 5-8 Myrs ago). The short-lived nature of the small and icy dust implies that the disruption occurred within the last year, or that replenishment due to ongoing collisional fragmentation is occurring.

Proposed Keywords : dusty disks, composition, spectroscopy, infrared, asteroids, comets

# 1. Introduction

One of the most dramatic Spitzer discoveries to date has been the extreme level of zodiacal emission around the nearby K0V star HD 69830 (Figure 1). The dust cloud around HD69830 has more than 1,000 times the emission of our own zodiacal cloud and shows a plethora of solid state features attributable to small, hot, crystalline silicate grains located within 0.5 - 1.0 AU of the parent star (Beichman *et al.* 2005; Figure 2). Such intense emission from dust in the inner solar system is exceedingly rare (Bryden *et al.* 2006; Beichman *et al.* 2006b). Interest in this cloud and its link to the evolution of planetary systems was heightened recently by the announcement that three Neptune mass planets orbit within 0.63 AU of the star (Lovis *et al.* 2006). With an estimated age of 3 - 10 Gyr, a near-solar primary ($T_* \sim 5385$ K, $L_* = 0.60$ $L_{solar}$, $M_* \sim 0.86$ $M_{solar}$, $R_* \sim 0.89$ $R_{solar}$), and a near solar metallicity ([Fe/H] = 0.015, Beichman *et al.* 2005; [Fe/H] = -0.05 ± 0.02, Lovis *et al.* 2006), the HD 69830 system promises to be an excellent laboratory for the study of the interaction been a debris disk and a solar-type planetary system. One of the major unanswered questions about this system is whether the dust seen by Spitzer comes from collisions within a particularly massive asteroid belt or from a swarm of comets released by planet-disk interactions, and whether it is transient or persistent in nature. Here we use the infrared mineralogical model derived from the Spitzer IRS observations of the Deep Impact experiment (A'Hearn *et al.* 2005, Lisse *et al.* 2006) to investigate the nature of the dust found around HD69830.

# 2. Data

The paper uses the low resolution (R~100) Spitzer Space Telescope (SST) InfraRed Spectrograph (IRS) (7-35 um) measurements previously described in Beichman et al (2005) (Figures 1 and 2). The reader is referred to that paper and to Beichman et al (2006a) for details of the data reduction. The separation of the emission due to the stellar photosphere from the "excess" due to the orbiting debris disk is estimated to be accurate to 1-2% of the photospheric level (Figure 1). As a test of the data reduction, we re-reduced the IRS data using the S13 pipeline from the Spitzer Science Center which uses a different flat fielding technique. After removing the stellar photosphere we compared the two different versions of the spectrum of the excess. The agreement was generally excellent: between the two spectra there are only 3 wavelengths out of 270 data points that deviated by more than 3σ from one another and only 15 points that deviated by more than 2σ. (The spectral modelling results are only different, within the errors of the fit, for the amorphous carbon smooth component which mimics the continuum at short wavelengths.) In the modelling discussed below we use the original published spectrum (Figure 2). The quoted uncertainties in the derived composition information (Table 1) are consistent with either reduced spectrum and the associated errors.

## 3. Modeling

Beichman *et al.* (2005) modeled the emission spectrum of the material in the debris disk by assuming a zeroth order model using dust with comet Hale-Bopp's optical parameters and 400 K temperature (Figure 2). From this, they deduced a total mass for the debris disk which exceeded the mass of any known comet suggesting an asteroidal rather than a cometary origin for the debris. However, the quality of their fit was less than ideal, as the nature of the dust in the two systems is significantly different upon detailed examination. In this work, a more sophisticated spectral modeling approach was taken, motivated by the successful analysis of the material excavated by the Deep Impact experiment from the interior of comet 9P/Tempel 1 (Lisse *et al.* 2006) and comet Hale-Bopp and YSO HD100546 (Lisse *et al.* 2007). As done for Tempel 1 and Hale-Bopp, the HD69830 excess flux was calculated by first removing the photospheric contribution using a Kurucz model fit, and then converting the remaining excess flux to an emissivity spectrum by dividing the measured fluxes by a best-fit blackbody (here found to be 400K from chi-squared minimization testing, in good agreement with the findings of Beichman *et al.* 2005). We work in emissivity space because it is very easy to pick out the contributions of the different mineral species once the gross effects of temperature are removed.

The resulting emissivity spectrum (Figure 3) was then compared to a linear combination of emission spectra of over 80 candidate mineral species (there are no obvious gas emission lines), selected for their reported presence in YSOs, solar system bodies, dusty disks systems, and IDPs. The results of the modeling are returned as the coefficients (weights) required to multiply each normalized mineral species emissivity spectrum in the best-fit linear sum spectrum (Figure 3 and Table 2). Assuming crystalline densities for the mainly small (~1 μm or less radius) dust, we can convert these weights, which are the observed surface area of each species referenced to a 400K blackbody at the distance of HD69830, into the relative number of moles of each species. To derive the total mass of each species, we integrate the best-fit PSD from 0.1 to 20 μm, the range of dust particle sizes the Spitzer spectrum is sensitive to, determining the absolute value by the amount of material required to produce the observed absolute flux.

When doing this comparison, the detailed properties of the emitting dust, i.e. the particle composition, size distribution, temperature, and porosity, all had to be addressed, as they can strongly affect the observed infrared flux. These quantities are thus products of the model solutions. The reader is referred to the discussion given at length in Lisse *et al.* (2006) and in the on-line supplemental material (SOM) for additional details of the calculations. Verification and validation of the model, as well as a measure of the range of modeling compositions found for cometary systems, can be found in the comparison of comets Hale-Bopp, Tempel 1, and the YSO HD100546 (Lisse *et al.* 2007).

We present here a model (Figures 3 and 4) consisting of the fewest and simplest dust species possible that result in a consistent fit to the observational data. A set of components was tested exhaustively before the addition of a new species was allowed, and only species that reduced the $\chi^2_\nu$ below the 95 % confidence limit (C.L.) were kept. It is important to note that we have not necessarily found the **exact** mineral species with our technique; rather, we claim to have found the important mineral classes present in the dust, and the gross atomic abundance ratios. It is also important to emphasize that, while the number of parameters (composition, temperature, particle size distribution (PSD)) may seem large, instead there are very few detected species for the 270 independent spectral points and 13 strong features obtained at high SNR by the SST/IRS over the 7-35 μm range. It was in fact extremely difficult to fit the observed spectrum within the 95% C.L. of $\chi^2_\nu$ = 1.16.

## 4. Derived Properties of the dust

**Temperature.** We modeled the dust excess around HD69830 both as an extended disk of dust and a localized dust torus. The toroidal models were motivated by the single temperature distributions found for many of the objects studied by Beichman *et al.* (2006a) and Chen *et al.* (2006) as well as the narrow dust structures found in many of the HST images of debris disks (Kalas *et al.* 2005, 2006). We find that only narrow toroidal models fit the data within the 95% C.L.. Our best-fit temperature of the smallest dust particles (0.1 - 1.0 μm), which superheat substantially above Local Thermal Equilibrium (LTE), is $T_{max}$ = 370 K for the olivines, 340 K for the pyroxenes, and 410 K for the amorphous carbon. The largest particles in our calculation, 1000 μm, are set to $T_{LTE}$, and dust of intermediate size is scaled between the two extremes (Lisse *et al.* 2006). Using the Tempel 1 ejecta temperatures as a guide (where we found olivines at 340K and amorphous carbon at 390K at 1.51 AU from the Sun, where LTE = 230K), and allowing for an HD69830 stellar radius that is 0.89 the Sun's, a luminosity that is 0.60 $L_{solar}$, and a $T_{eff}$ that is 5385 K (Lovis *et al.* 2006), we estimate a location for the dust at 0.93 - 1.16 AU from HD69830. This range of distances is coincident with the strong 2:1 and 5:2 mean motion resonances of HD69830d predicted by Lovis *et al.* (2006) from their Monte Carlo modeling of stable dust particle orbits, and the inner edge of a large distribution of low eccentricity stable orbits (Figure 5).

Interestingly, the best-fit single continuum temperature to the 7 - 35 um Spitzer spectrum is 400K, close to the amorphous carbon temperature. As seen in the Tempel 1 ejecta from the Deep Impact experiment (Lisse *et al.* 2006), the amorphous carbon dominates the continuum behavior as it is the most continuum-like and the hottest material, thus contributing most to the short-wavelength emission. (For the Tempel 1 ejecta, at 1.51 AU from the Sun, the amorphous carbon temperature for the smallest, dominant dust was 390K, while the olivines and pyroxenes reached about 340K, and the water ice about 220K.) It is the short wavelength emission which is most diagnostic of an SED's continuum temperature, because of the steep falloff off the blackbody on the Wien law side of the emission peak and the relatively shallow, and temperature independent behavior on the long wavelength Rayleigh-Jean side of the peak.

For scale, putting the dust at the mean distance of HD69830d, or 0.63 AU from the central source, and assuming greybody behavior for the dust (i.e., ignoring any superheating due to differences in the particle optical and IR emissivities) yields dust at $T_{LTE} = 245/\sqrt{(0.63 \text{ AU})} = 309K$; dust at the location of HD69830c would have $T_{LTE} = 245/\sqrt{(0.19 \text{ AU})} = 562K$; dust at 1 AU, the 2:1 mean motion resonance for the outermost Lovis planet (HD69830d), would have $T_{LTE} = 245K$ (Figure 5). Thus, had we not done the detailed compositional modeling and assumed instead that the 400K continuum temperature was equal to $T_{LTE}$, we would have placed the dust erroneously at a distance of approximately 0.4 AU from the central source. The presence of abundant water ice, though, in our best-fit compositional solution is a clue to that this would be incorrect, as water ice would be stable for only seconds at 0.4 AU.

Extrapolating the best-fit model spectrum longwards of 35 um using an assumed absorption efficiency $Q_{abs} = Q_{abs}(35 \mu m/\lambda)$, where $\lambda$ is the wavelength, as a lower limit and $Q_{abs}$ = constant as an upper limit, we estimate a 70 um flux of 5 ± 4 mJy, while the Spitzer MIPS measurements of Beichman *et al.* (2005) yield a value of 7 ± 3 mJy for the 70 um excess. We thus concur with conclusion that there is no large amount of cold (i.e., Kuiper Belt dust) contributing to the excess emission detected by Spitzer, and that the emission is dominated by the single debris disk at T~ 400 K .

**Size Distribution and Total Mass.** The best-fit Particle Size Distribution (PSD) ~ $a^{-3.9\pm0.2}$ used for all species argues for dust dominated by small particles in its emitting surface area, and by particles of all sizes in its mass. The predominance of small particle surface area is why we see the strong emission features  - the dominant particles are optically thin, and the feature to continuum ratio is high. This very steep PSD  is unusual; more gradual ones are typically found for dusty disks systems. On the other hand, a system in collisional equilibrium would demonstrate a PSD ~ $a^{-3.5}$ (Dohnanyi 1969, Durda and Dermott 1997); for "real" systems a size distribution even steeper than $a^{-3.5}$ at small sizes is expected in a collisional cascade both because of its truncation at the blow-out limit (Thebault, Augereau and Beust 2003) and because of the dependence of particle strength on size (O'Brien and Greenberg 2003).

Integrating the mass implied by the best-fit PSD, we find an estimated dust mass of ~3 x $10^{17}$ kg **definitively** detected by the Spitzer IRS spectrum (i.e., mass in particles of 0.1 - 10 μm in size that contribute appreciably to the $\chi^2_\nu$ value of the fit). However, the majority of the mass is in the largest, optically thick particles if they are present (dm/dloga ~ $a^0$). While the Spitzer IRS measurements are not very sensitive to these large grains, the amount of mass in the larger (20 μm  - 10m, Beichman *et al.* 2005) can be estimated by assuming a PSD of constant slope, and summing the mass of all particles up to a cutoff of 10 m. In this case we find a total disk mass of 1.8 x $10^{18}$  kg, or 3 x $10^{-7}$ $M_{Earth}$. We find a total estimated 0.1 - 10 μm  surface area of 1.2 x $10^{20}$ $m^2$. Assuming the dust is in a ring centered at 1.0 AU from the central star (Figure 5), in an annulus 0.1 AU wide and with total surface area 1.4 x $10^{21}$ $m^2$,  this implies a spatial filling factor for the dust of 0.9 x $10^{-2}$. In comparison, the coma of a bright, active comet, like Halley or Hale-Bopp, is observed to have an average spatial filling factor of the coma dust in the central $10^5$ km on the order of $10^{-7}$ (Lisse *et al.* 1998, 2004). It is only in the collisionally thick parts of the coma, within 10 km of the nucleus, where the coma dust spatial filling factor reaches values $\geq 10^{-2}$.

**Compositional and Atomic Abundance.** The derived compositional abundances from our best-fit dust model, with $\chi^2_\nu = 1.09$, are given in Table 1. As a test of the robustness of our best fit model, the last column in Table 1 also gives the best-fitting $\chi^2_\nu$ found after deleting a particular species from the compositional mix and re-fitting the data. Unlike the case of Tempel 1 and Hale-Bopp, there is no definitive evidence for water gas, PAHs, phyllosilcates, or sulfides (Figures 3 and 4). The relative atomic abundance we find in the circumstellar dust is H:C:O:Si:Mg:Fe:S:Ca:Al = 0.35 : 0.90 : 4.3 : 1.0 : 1.2 : 0.4 : 0.0 : 0.06 : 0.0 (for Si = 1.0). Unlike the case of Tempel 1, Hale-Bopp, and HD100546, the Si:Mg:Fe:Ca:Al refractory element ratios are inconsistent with solar system abundances determined from the Sun and from C1 chondrites (Anders and Grevesse 1989), with Fe, S, and Al being highly deficient (Figure 6).

**Detailed Mineralogy.** Here we discuss a few of the standout materials for the case of HD69830. We refer the reader to Figures 3 and 4 while reading this section.

**Silicates.** We find a best-fit model with a spectrum dominated by emission from silicates, predominantly Mg-rich olivines (80%) and crystalline pyroxenes (20%) (Figure 3). Unlike our results on cometary systems (Lisse *et al.* 2006, 2007), amorphous pyroxene is conspicuously lacking. The ratio of total olivine to pyroxene abundance is ~ 4:1, as compared to the ratios of ~1:1 in the Tempel 1 sub-surface ejecta (Lisse *et al.* 2006), ~2:1 in the dust emitted into the coma of comet C/1995 O1 (Hale-Bopp), and 1:1 in the YSO HD100546 (Lisse *et al.* 2007). Both findings are direct evidence for high temperature processing (greater than the 900K annealing temperature of pyroxenes) of the rock forming species. The relative lack of pyroxenes is what causes the deficiency in the HD69830 spectrum vs. the Hale-Bopp spectrum at 8 - 10 μm (Figure 2). The crystalline pyroxenes that are present contain a broad range of mineralogical components, including diopside, ferrosilite, and highly processed bronzite, but not simple ortho- or clino-pyroxene. Two different kinds of Mg-rich olivine were needed to accurately fit the spectrum, one of which has not been found in cometary systems, testimony to the super-abundance of olivine in the dust. There is a significant fraction of amorphous olivine detected, ~25% of all olivine, arguing that some of the rock forming elements were not raised to temperatures of 1200K and above for significant amounts of time.

**Iron and Sulfur/Sulfides.** Some of the major sources of iron found in the cometary systems, the carbonate siderite ($FeCO_3$) and the Fe sulfides (pyrrohtite, ningerite, and pentdantalite), are totally lacking in the HD69830 debris disk. Iron is only obviously present in the minority iron-bearing silicates, and is clearly under-abundant vs. solar and C1 abundances, and the other elements (Figure 5). There is no obvious source of S in the debris disk spectrum.

**Carbon Bearing Species.** The major sources of carbon in the comet systems, amorphous carbon and PAHs, are lacking in the HD69830 best fit model to the spectrum published in Beichman *et al.* (2005). The only carbon clearly present is in the form of carbonate, and is clearly under-abundant vs. solar and C1 abundances, and the other elements. Even without 5 - 8 μm spectral coverage, it is clear that the 8.6, 11.3, and 12.5 μm PAH emission lines are lacking at the spectral sensitivity and resolution of the IRS low resolution modules. Higher resolution observations with IRS are planned that might reveal a low level of PAH emission.

This presence or absence of amorphous carbon is important in that except for this component, there is a good match, including the carbonates, between the best fit compositional model and the composition of outer main belt asteroids. As the amorphous carbon signature is similar to a smooth continuum, it is possible that some of the emission from amorphous carbon has been misinterpreted as stellar photosphere emission, and used to set the stellar contribution baseline. As there is currently some uncertainty in the removal of the stellar photospheric contribution which is critical to the detection of amorphous carbon, we quote here an upper limit for the amount of amorphous carbon in the debris disk by moving the photospheric model to the lowest possible allowed 2-sigma limit (2%) below the nominal normalization. Doing so, we find the exact same mix of materials but with detection of amorphous carbon. We conclude that the maximum amount of amorphous carbon possible in the system to be a significant 1.8 moles (relative), or about 38% by mole fraction of the total material, 6% by surface area. Carbon at this level of abundance is very typical of that seen in the cometary bodies (Lisse *et al.* 2006, 2007). The minimum possible amount of amorphous carbon present is zero.

**Water ice**. The clear presence of water ice at 11 to 15 um is somewhat surprising at first look, given that the smallest dust particles appear to be as hot as 400K, and the average dust particle temperature is ~340 K, much above the ~200K temperature at which water ice freely sublimates at 0 torr. However, $T_{LTE}$ at 1 AU is 245K, still somewhat above the ice stability temperature, but close enough to it that cooling by evaporative sublimation can stabilize the ice if it is isolated from the other hot dust. We saw the same large dichotomy in dust temperatures for the Tempel 1 ejecta.

If the water ice is unassociated with the debris disk at 1 AU, but is instead present in the "Kuiper Belt" for the system and is serendipitously along the line of sight, then we expect long-lived ice without much water gas formed by evaporation. However, significant icy dust in a Kuiper Belt at ~50 AU from the star is in direct contradiction to the 3σ upper limit on the 70 um flux observed for the system by Beichman *et al.* (2005). Although water ice has a strong emission feature at 65 um the lack of 70 um emission is consistent with a simple extrapolation of emission from hot, small grains and the absence of any colder (larger) grains.

Unfortunately, we do not have 5 - 8 μm measurements of the HD69830 system in hand at this time. This region is highly diagnostic not only of water gas but also of carbonate and PAH materials found in the comets and HD100546 (Lisse *et al.* 2007). Without 5 - 8 μm spectroscopy, we cannot determine the presence or absence of related water gas in the system, but we hypothesize from our knowledge of water ice's thermal stability that there must be an abundance of water gas as well. On the other hand, we can say that water ice is not super-abundant. We do know that the main emission from water ice is at 11 - 15 μm, and the superabundance of water ice particles in the ISO Hale-Bopp spectrum is what caused the mismatch seen in this region in the Beichman *et al.* 2005 dust model.

## 5. Results and Discussion

Taken together, the compositional and physical nature of the dust determined from the Spitzer spectrum provides strong evidence for differentiation and processing of material in ongoing near-collisional equilibrium. This disk is derived from material that is not primordial. Here we discuss further the implications for the putative parent body source of the dust.

**Dynamical Considerations.** The derived HD69830 minimum dust mass of $3 \times 10^{17}$ kg is equivalent to the mass of a very large 60 km radius comet, assuming a density of 0.35 g cm$^{-3}$ (from Deep Impact; A'Hearn *et al.* 2005, Richardson *et al.* 2007). While larger particles may be present, we cannot prove this with the IRS measurement alone. Sensitive mm/submm observations (with the JCMT-SCUBA2, for example) would help us to understand the larger particle sizes better. The critical point here, however, is that we cannot **definitively** say the mass of material is much larger than that of a comet in our solar system - comet Hale-Bopp may have been of this size. If the body were an asteroid, instead, with density 2.5 g cm$^{-3}$, then the minimum radius required (assuming total disruption) would be 30 km. We can say that an asteroidal body of such size is found very commonly in the solar system.

The observed dust must be relatively fresh. Due to the effects of solar radiation pressure, the lifetime for dust of radius $\leq 1$ μm (particles with $\beta = F_{rad\ press}/F_{grav} > 0.5$; Finson and Probstein 1968, Lisse *et al.* 1998) in the solar system in the vicinity of 1.0 AU can be measured in weeks to months. The situation for the near-solar K0V HD69830 system will be little different. The lifetime for dust spiraling into the Sun due to Poynting-Robertson (PR) drag is given by $\tau_{PR}$ =$10^5$ yrs * a(μm) (Burns *et al.* 1979), so even dust as large as 1 cm is cleared out within 1 Gyr. With a system age of 3 - 10 Gyrs, the observed excess cannot be due to primordial dust created during the formation of the system. The presence of significant quantities of short-lived water ice (lifetimes of seconds to months vs. evaporation inside the HD69830 iceline at 1.5 AU) also argues for recent generation of these small grains.

New work by Wyatt *et al.* (2007) also demonstrates that it is highly unlikely that an asteroid belt formed from an original nebula could create the observed excess at 3 Gyr in steady state, as collisional processing would have efficiently removed most of the mass of this belt over such a timescale. The survey of Bryden *et al.* (2006) shows that sun-like stars which, like HD69830, have hot dust at a few AU are rare, found around just 2% of stars. Similar results were found by Rieke et al. (2005) and Song et al. (2005). It is considered that the majority of such systems are also unlikely to be explained as massive asteroid belts (Wyatt et al. 2007). Thus studies of the aggregate population of dusty disks also argue for an extremely recent, stochastic formation of the HD69830 dust complex.

**Solar System Analogue for the Parent Body.** Given that the likelihood of serendipitously observing the breakup of an asteroid within weeks of its occurrence is unlikely, and that Mannings and Barlow (1998) found a 25 um excess for the system in 1983, we conclude that the small particle and icy dust components must be undergoing continual replenishment on timescales of decades or longer. The most likely physical source of continuous replenishment is a dense, localized debris belt undergoing frequent collisions between rapidly moving fragments of size 100 μm - 10 m, forming fine dust in a manner similar to the zodiacal bands in our solar system. In the case of our zodiacal bands, such collisions, occurring 5 to 8 Myrs ago, resulted from the collisional formation of the Karin and Veritas family of asteroids from larger parent bodies (Nesvorny 2006). The sheparding action of the outermost Neptune-sized planet in the system may also be important in the concentration and dynamical stirring of the debris fragments. By far the largest range of eccentric particle orbits are stably bound in the 2:1 resonance at 1 AU, where we place the dust from our modeling, than in any other resonance for the system (Lovis *et al.* 2006). The large range in eccentricities will cause more frequent and violent collisions for this resonance, as bodies captured into the resonance on very different orbits intersect. We thus find the collisional disruption of a body in the resonance and the localization of the collision products into the 2:1 mean motion resonance (Figure 5) very plausible.

Given the compositional results of our modeling for the majority species constituting the dust in the debris disk around HD69830, and using the small bodies in our solar system as a guide, we find the best match to the compositional model derived here to be a P or D-type asteroid (in the Tholen Taxonomy, c.f. Lodders and Fegley 1998) with low carbon abundance. Comets and their immediate ancestors the Centaurs and KBOs are ruled out due to the lack of amorphous carbon, sulfides and amorphous pyroxene (Zolesnksy *et al.* 2006, Lisse *et al.* 2006, 2007). S-type asteroids, also abundant in the inner main belt, do not contain water or products of aqueous alteration like carbonates. C-type asteroids contain much more carbon and phyllosilicates than found for the HD69830 dust.

The P/D-type asteroids represent the most numerous asteroidal bodies in the outer main belt and in the Jovian Trojan swarms, and a typical body is of 20 to 100 km radius, so it is entirely plausible that the mass of debris measured for the HD69830 system could have come from their disruption, even for the larger size estimates of the original parent body determined by Beichman *et al.* (2005). Also, the recent finding of "comet-like asteroids" by Hseih and Jewitt (2006) in the solar system, spectroscopic evidence for abundant hydrogen in asteroids (Rivkin *et al.* 2004), and the evidence for aqueous alteration on Ceres (Russell *et al.* 2004, McCord and Sotin 2005) makes the finding of appreciable water ice produced by ongoing collisional fragmentation of a P or D-asteroid credible.

The one major issue with this determination is the location of formation of the P/D-type asteroids - they represent the majority species in the outer main belt of our solar system, at 3 - 5 AU from the Sun today, outside the current ice line, and were presumably formed where conditions were cold enough to trap and hold most of the volatile water present. To have the dust from such a body at ~1 AU in HD69830, we must also posit a mechanism for moving the body inwards after formation, e.g. planetary migration (Alibert *et al.* 2006). This mechanism may also be responsible for moving the Neptune-sized bodies inwards as well. Analogous objects are the mainbelt asteroids in the 3:2 and 4:3 resonances with Jupiter. Another possibility is an orbital perturbation large enough to cause an icy parent body asteroid to be captured into the 2:1 or 5:2 HD69830d resonances. Analogous objects are the JFC comets in the 1:2 resonances with Jupiter. The possibility of other, smaller planets in the system acting as efficient scatters of icy bodies in the outer reaches of HD69830, is high, given the presence of 3 Neptune-sized planets, but until said planets are located, the size and nature of their scattering cannot be estimated.

## 6.  Conclusions

Starting from a knowledge of small body mineralogy derived from the recent Deep Impact experiment excavation of comet 9P/Tempel 1, we are able to easily fit the observed disk excess emission observed by Beichman et al. using the Spitzer Space Telescope IRS. From this fitting, we find that

- The IR excess is due to material in a dense disk orbiting around the star at $1.04 \pm 0.12$

AU, coincident with the 2:1 and 5:2 orbital resonance with the outermost Neptune sized planet, HD69830d, and other stable low eccentricity orbits with respect to the entire system.

- The dust is totally lacking in amorphous pyroxenes PAHs, and metallic sulfides, and is deficient in Fe and S vs. solar abundances.

- The material is experiencing ongoing collisional fragmentation, as evidenced by the continual supply of ephemeral small and icy dust particles.

- The material originated in a parent P- or D-type asteroidal body originating most likely in an asteroid belt in the 2:1 resonance. The breakup of this body appears analogous to the fragmentation that accompanied the formation of the Karin and Veritas families in the solar system.

## 7. Acknowledgements

This paper was based on observations taken with the NASA Spitzer Space Telescope, operated by JPL/CalTech, and modeling performed under JPL contract 1274485. The authors would like to thank A. Cheng, O. Barnouin-Jha, C. Lovis, A. Rivkin, and A. Roberge for valuable input, and R. Hurt for help with the figure graphics.

## 9. Tables

**Table 1 - Composition of the Best-Fit Model to the SST IRS HD69830 Spectrum**

| Species | Weighted[a] Surface Area | Density (g cm$^{-3}$) | M.W. | $N_{moles}$[b] (relative) | Model $T_{max}$ (°K) | Model $\chi^2_\nu$[c] if not included |
|---|---|---|---|---|---|---|
| **Detections** | | | | | | |
| *Olivines* | | | | | | |
| Amorph Olivine (MgFeSiO$_4$) | 0.19 | 3.6 | 172 | 0.40 | 370 | 13.3 |
| ForsteriteKoike (Mg$_2$SiO$_4$) | 0.20 | 3.2 | 140 | 0.45 | 370 | 4.83 |
| Forsterite038 (Mg$_2$SiO$_4$)[d] | 0.33 | 3.2 | 140 | 0.74 | 370 | 15.9 |
| Fayalite (Fe$_2$SiO$_4$) | 0.04 | 4.3 | 204 | 0.07 | 370 | 1.29 |
| *Pyroxenes* | | | | | | |
| Bronzite (Mg$_{1-x}$Fe$_x$Si$_2$O$_6$)[d] | 0.13 | 3.5 | 232 | 0.19 | 340 | 4.47 |
| FerroSilite (Fe$_2$Si$_2$O$_6$) | 0.09 | 4.0 | 264 | 0.14 | 340 | 2.46 |
| Diopside (CaMgSi$_2$O$_6$) | 0.05 | 3.3 | 216 | 0.075 | 340 | 1.53 |
| *Carbonates* | | | | | | |
| Magnesite (MgCO$_3$) | 0.07 | 3.1 | 84 | 0.24 | 370 | 1.54 |
| Dolomite (CaMgC$_2$O$_6$)[d] | 0.04 | 2.9 | 184 | 0.064 | 370 | 1.25 |
| Water Ice (H$_2$O) | 0.08 | 1.0 | 18 | 0.44 | 220 | 2.28 |
| Amorphous Carbon (C) | 0.07 | 2.5 | 12 | $\leq 1.5$ | 410 | 1.78 |
| **Upper Limits** | | | | | | |
| Sulfides (Mg$_{50}$Fe$_{50}$S) | 0.03 | 4.5 | 72 | $\leq 0.02$ | 370 | 1.22 |
| PAH (C$_{10}$H$_{14}$),ionized | 0.05 | 1.0 | <178> | $\leq 0.028$ | N/A | 1.10 |

(a) - Weight of the emissivity spectrum of each dust species required to match the HD69830 emissivity spectrum.
(b) - $N_{moles}(i) \sim$ Density(i)/Molecular Weight(i) * Normalized Surface Area (i). Errors are ± 10% (2σ).
(c) - Total best fit model $\chi^2_\nu = 1.09$.
(d) - Not found in cometary systems to date.

## 10. Figures

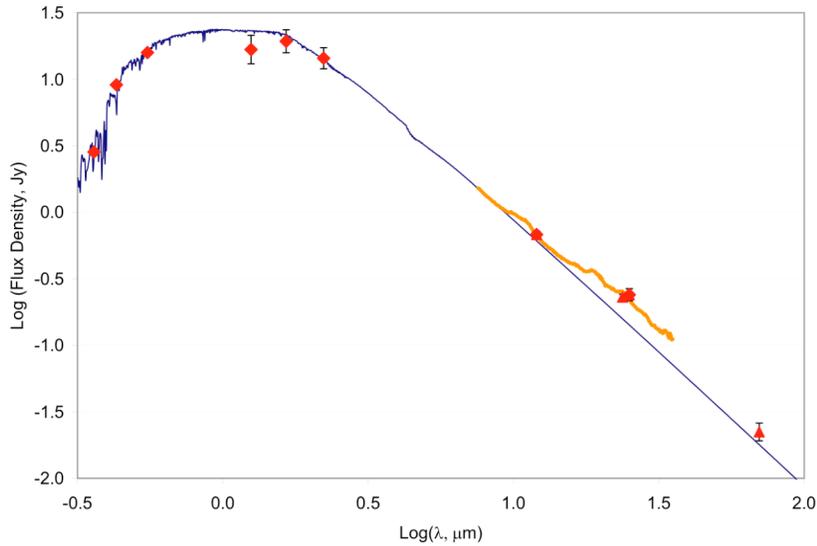

*Figure 1* - **SED of HD 69830** derived from *Hipparcos* measurements (optical photometric points), 2MASS (*J*, *H*, and *Ks photometric points*), Spitzer 7 - 35 um spectroscopy (green curve), IRAS 12 and 25 um photometry (Mannings and Barlow 1998), and Spitzer 24 and 70 um photometry. Error bars are 1σ, and are smaller than the symbol size for some points. All Spitzer data is from Beichman *et al.* 2005. Overlaid is the Kurucz model (*blue*) fit to the short wavelength photometry used to remove the stellar photospheric contribution. The 2MASS data are very uncertain because of the brightness of the star.

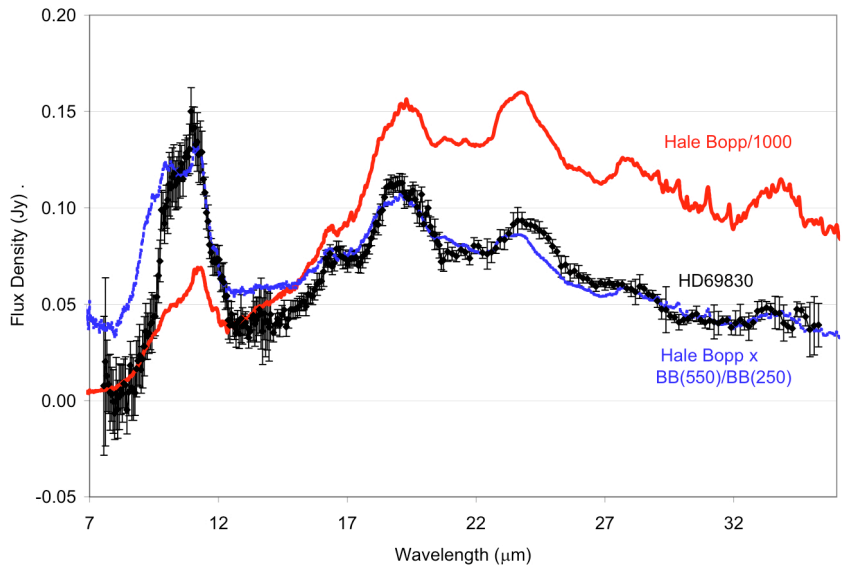

*Figure 2* - **Comparison of the HD69830 and Hale-Bopp emission spectra.** Black - Spectrum of the excess of HD 69830 according to Beichman *et al.* 2005. Red - For comparison, the spectrum of Comet C/1995 O1 (Hale-Bopp), following Crovisier *et al.* 1997. Blue - Hale-Bopp spectrum normalized to 550K to best match the HD69830 spectrum. Note the comparative excess of emission for Hale-Bopp at 8 - 10 um due to pyroxene, and at 12 - 15 um due to water ice.

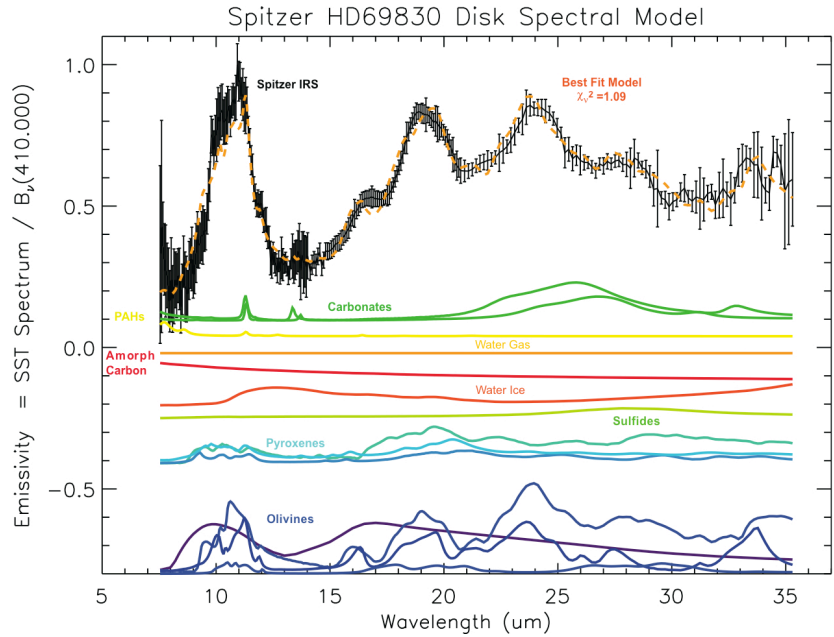

*Figure 3 -* **Emissivity spectrum for HD69830 excess, as measured by the Spitzer IRS.** The data and model shown are for the smallest photospheric subtraction, demonstrating the upper limits to the abundance of PAHs, amorphous carbon, and sulfides in the system. Error bars are ±2σ. Black : SST excess spectrum, divided by a 400K blackbody. Orange dashed line: best fit model spectrum. Colored curves: emission spectra for the constituent species, scaled by the ratio $B_\lambda(T_{dust}(a)_i)/B_\lambda(T_{continuum})$, with $T_{continuum}$ = 400K. The best-fit model individual species curves have been scaled by a factor of two versus the data for emphasis. Purples - amorphous silicates of pyroxene or olivine composition. Light blues - crystalline pyroxenes: ferrosilite, diopside, and orthoenstatite, in order of 20 μm amplitude. Dark blues - crystalline olivines, forsterite and fayalite, in order of 20 μm amplitude. Red - amorphous carbon. Deep orange - water ice. Light orange - water gas. Yellow - PAHs. Bright greens - carbonates: siderite and magnesite, by order of 7 μm emissivity amplitude. Olive green - sulfides.

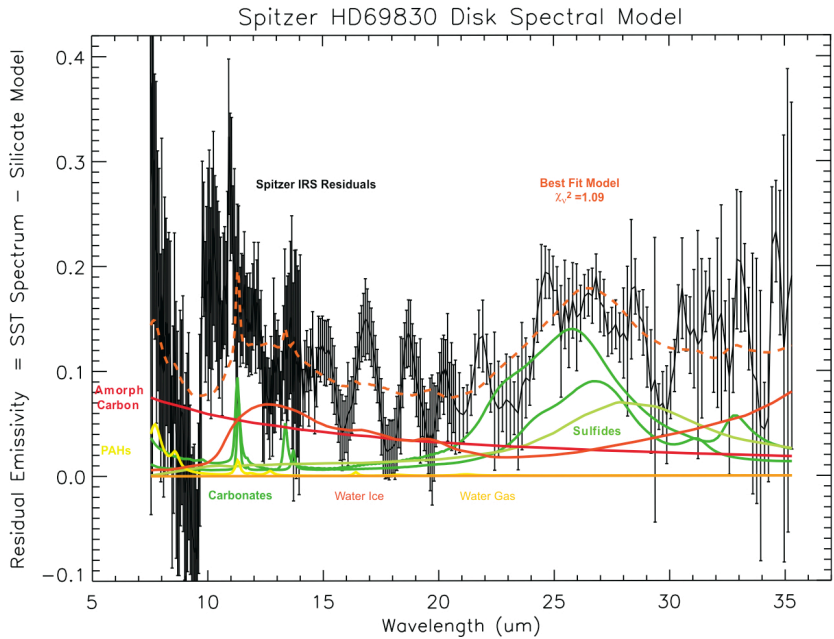

*Figure 4 -* **Spectrum after subtraction of the best-fit silicate components**, the dominant species in the emission. Scale, color and labels are the same as in Figure 3. The sinusoidal features at 15 - 20 μm are artifacts ('fringes') of the IRS instrument. The true signal is near the average of the excursions.

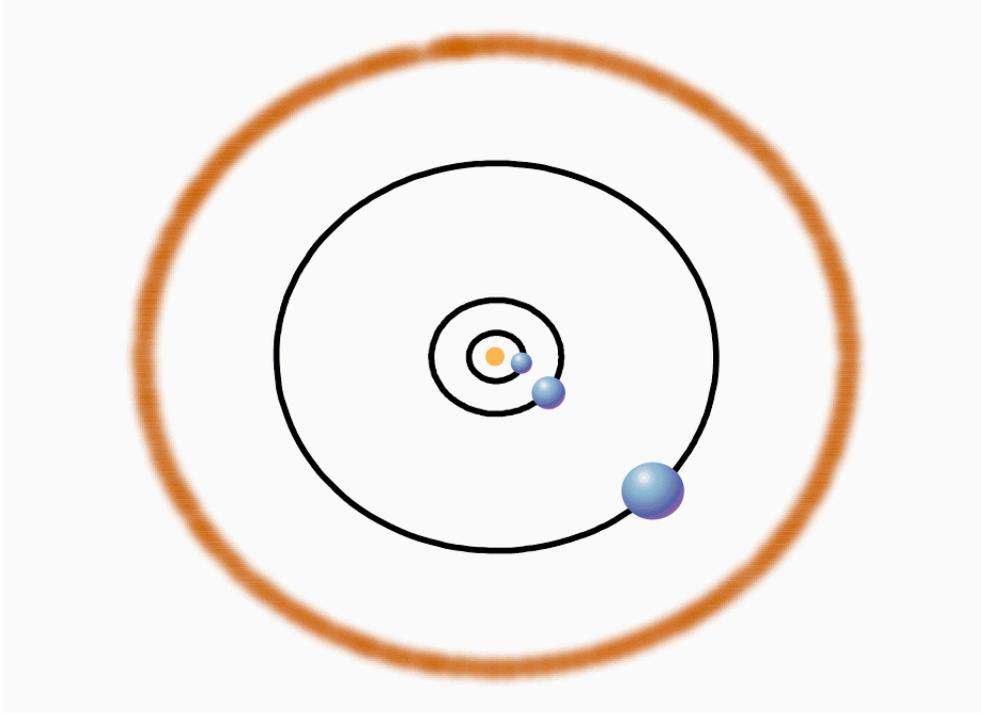

*Figure 5* - **a) Schematic of the HD69830 system, as it is understood from the present work and that of Lovis et al. (2006).** In order from the center of the system : Orange-yellow center filled circle - the HD69830a K0V primary, at $M_* \sim 0.86\ M_{solar}$, $R_* \sim 0.89\ R_{solar}$, $T_* \sim 5385$ K, $L_* = 0.60\ L_{solar}$,; filled blue circles - HD69830b at 0.0785 AU and $M \geq 10.2\ M_{Earth}$, HD69830c at 0.186 AU and $M \geq 11.8\ M_{Earth}$, and HD69830d at 0.63 AU and $M \geq 18.1\ M_{Earth}$; **fuzzy brown curve** - the thick dust belt giving rise to the IR excess detected by Spitzer, located at $\sim 1$ AU, coincident with the 2:1 resonance of HD69830d, and of total mass $\geq 3 \times 10^{17}$ kg. *N.B.* - The planets are shown as following idealized simple circular trajectories. **b) Stability analysis for massless particles in the HD69830 system, after Lovis *et al.* (2006).** The color grid corresponds to this stability index, red denoting the most unstable orbits with a close encounter with the central star or a planet, while dark blue corresponds to very stable orbits. The particles can survive for an extended time in the stable region beyond 0.8 AU. As for the Solar System main asteroid belt, several mean motion resonances with the outermost planet can be identified that would create gaps or accumulation in a potential asteroid belt: 1:2 (,0.40 AU), 2:3 (~0.48 AU), 1:1 (~0.63 AU), 3:2 (~0.82 AU) , 2:1 (~1.00 AU), and 5:2 (~1.18 AU).

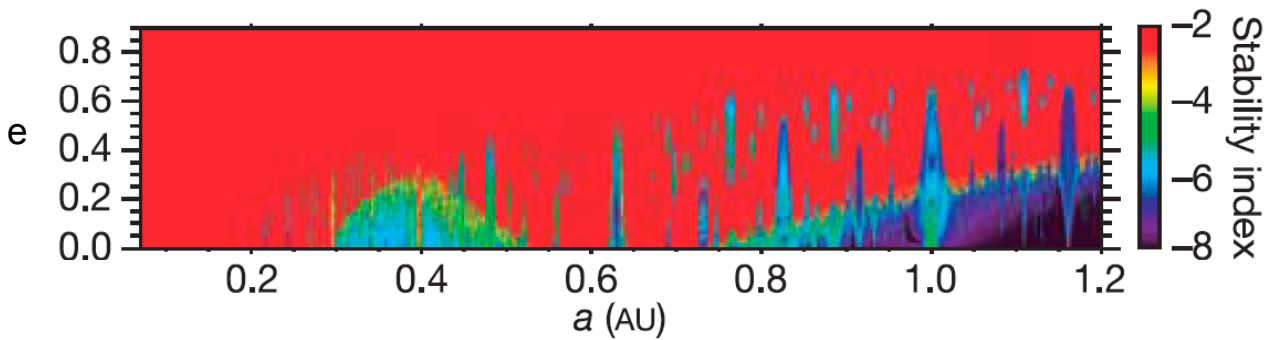

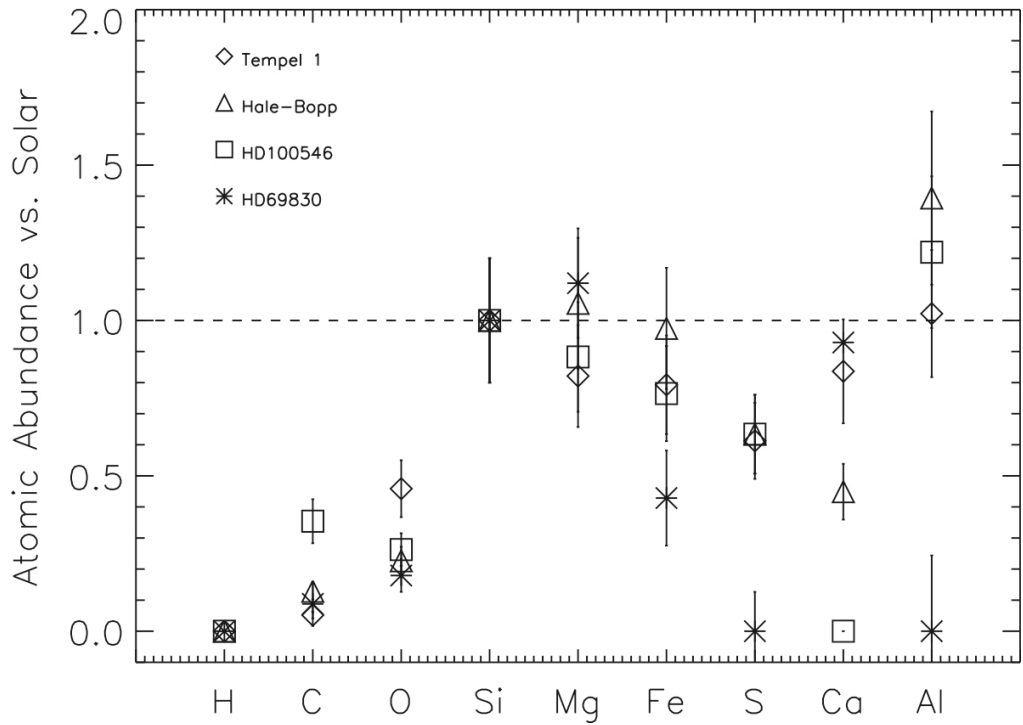

*Figure 6 -* **Atomic abundances for the circumstellar dust around HD69830, compared to the 3 comet systems studied to date with the Tempel 1 ejecta compositional model.** Abundances are given vs. solar, assuming Si = 1.0 Error bars for the relative measures are ± 20% (2σ). Diamonds = Tempel 1, Squares = HD100546, Triangles = Hale-Bopp, Stars = HD69830. The nominal solar value is denoted by the dashed line. HD69830 is markedly different than the range of cometary abundances in Fe, S, and Al. *N.B.* - the C abundances for Hale-Bopp and HD100546 may be artificially high due to the ISO-SWS calibration (see Lisse *et al.* 2007), while Tempel 1 hydrogen and oxygen abundances includes the contribution of $H_2O$ gas. The significant cometary reservoirs of gas phase CO and $CO_2$ are not accounted for here.